\documentclass[aps,prl,twocolumn,superscriptaddress]{revtex4-1}

\usepackage{graphicx}
\usepackage{amsmath}

\begin{document}


\title{Isotropic Topological Second-Order Spatial Differentiator \\ Operating in Transmission Mode}


\author{Olivia Y. Long}
\affiliation{Department of Applied Physics, Stanford University, Stanford, California 94305, USA}
\author{Cheng Guo}
\affiliation{Department of Applied Physics, Stanford University, Stanford, California 94305, USA}
\author{Haiwen Wang}
\affiliation{Department of Applied Physics, Stanford University, Stanford, California 94305, USA}
\author{Shanhui Fan}
\email[]{shanhui@stanford.edu}
\affiliation{Department of Applied Physics, Stanford University, Stanford, California 94305, USA}
\affiliation{Ginzton Laboratory and Department of Electrical Engineering,
Stanford University, Stanford, California 94305, USA}


\date{\today}

\begin{abstract}
Differentiation has widespread applications, particularly in image processing for edge detection. Significant advances have been made 
in using nanophotonic structures and metamaterials to perform such operations. In particular, a recent work demonstrated a topological differentiator in which the transfer function exhibits a topological charge, making the differentiation operation robust to variation of operating conditions. The demonstrated topological differentiator, however, operates in the reflection mode at off-normal incidence and is difficult to integrate into compact imaging systems. In this work, we design a topological differentiator that operates isotropically in transmission mode at normal incidence. The device exhibits an optical transfer function with a symmetry-protected, topological charge of $\pm 2$ and performs second-order differentiation. Our work points to the potential of harnessing topological concepts for optical computing applications.
\end{abstract}


\maketitle

\section{Introduction}

Differentiation is a keystone mathematical operation that is especially important in image processing, where spatial differentiation is widely used for edge detection \cite{gonzalez_digital_image_book}. Traditionally, image processing has been implemented using digital computing, which suffers from low energy efficiency and slow speed due to the analog-to-digital conversions and discretization process \cite{zangeneh_review_2020}.
In contrast, optical analog computing offers the advantages of high-speed operation, parallel processing, and low power consumption \cite{zangeneh_review_2020,solli_review_2015}. 
In the conventional Fourier optics approach to analog computing, 4F correlator optical systems are used manipulate incident fields in Fourier space \cite{goodman, Abdollahramezani_2020_review_bulkyfourier}. However, such setups are bulky and difficult to integrate into imaging devices, motivating the search for alternatives. 

Following the first demonstration of analog processing using metamaterials \cite{Silva_2014}, there has been great interest in employing such designs to miniaturize bulky optical elements \cite{Rajabalipanah_program_metasurf_2021, Zhou_analog_metasurf_2020, Chizari_analog_comp_2016, dong_HCG_diff_APL_2018, Wan_dielec_metasurf_diff_2020, cordaro_metasurf_2019, pors2015analog, zangeneh_2019_topological}. To date, 
spatial differentiation has been achieved using phase-shifted Bragg gratings \cite{doskolovich2014spatial, Bykov_Laplace_2014}, surface plasmonics \cite{zhu_plasmonic_2017, Ruan_SPP_2015, hwang2018plasmonic}, the spin Hall effect of light \cite{tengfeng_spinhall_2019, he_geom_SHE_2020}, and photonic crystals \cite{Cheng_2018_optica, wang2020compact}. Other differentiators include Pancharatnam-Berry phase gradient metasurfaces \cite{Zhou_PNAS_2019}, arrays of split-ring resonators \cite{kwon2018}, and high-contrast gratings \cite{dong_HCG_diff_APL_2018}.
However, these structures typically rely on resonant modes to achieve differentiation. This results in inherently narrow bandwidths and high sensitivity to changes in operating conditions \cite{Cheng_2018_optica,kwon2020,zhou2020flat,Wan_dielec_metasurf_diff_2020, cordaro_metasurf_2019, komar2021edge}.
Moreover, differentiation is often anisotropic due to the lack of symmetry in the structure \cite{kwon2020,zhou2020flat,Wan_dielec_metasurf_diff_2020, kwon2018}.

Recently, a topological differentiator based on reflection from a dielectric interface was demonstrated \cite{tengfeng_2021}. In contrast to previous designs, this differentiator exhibits a nontrivial topological charge in the transfer function, making it more robust against geometric imperfections or environmental perturbations \cite{mermin1979topological}. Isotropic differentiation is achieved by selecting the appropriate input and output polarizations, as well as the incident angle. However, this structure operates in reflection mode at oblique incidence, which hampers direct integration into compact imaging systems. Moreover, such a device can only achieve first-order differentiation, which is less effective for image sharpening compared to higher-order derivatives \cite{gonzalez_digital_image_book}.

In this work, we provide further improvement of the concept of topological differentiator by demonstrating isotropic differentiation via transmission through a thin-film structure at normal incidence. Using circularly polarized incident light, 
we achieve second-order differentiation by transmitting only the orthogonal polarization.
Thus, our structure is superior to first-order differentiators in enhancing fine detail \cite{gonzalez_digital_image_book}. Moreover, the transfer function of this setup exhibits a topological charge of $\pm 2$, in contrast to previous designs with either no charge or $\pm 1$ charge \cite{tengfeng_2021}. This charge is further protected by the rotational symmetry of our structure. 
Through numerical demonstration, we show that edge detection can be achieved with high resolution and broadband operation.

\section{Results}
\begin{figure}[ht!]
\centering
\includegraphics[width=\linewidth]{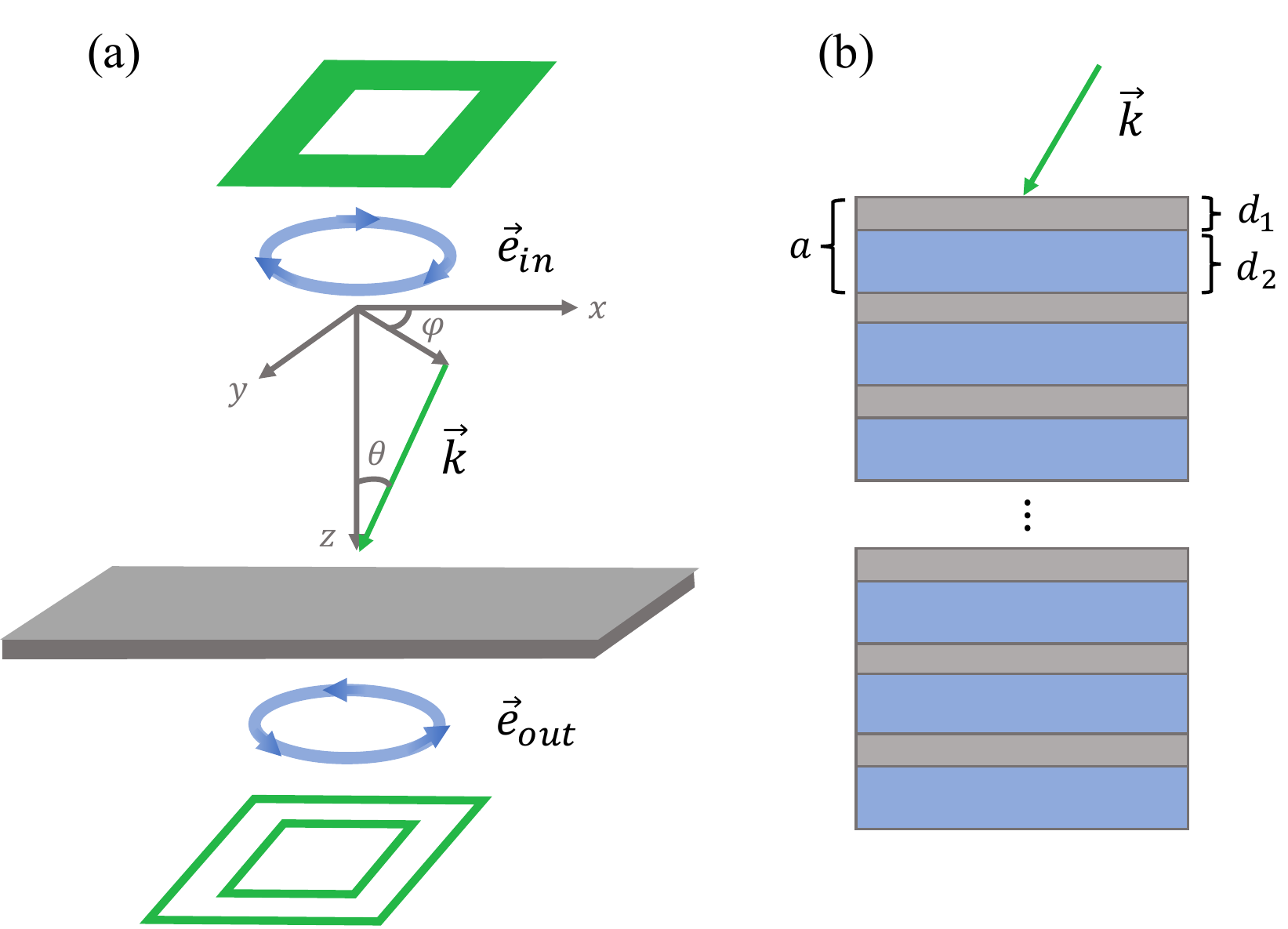}
\caption{(a) Schematic of isotropic topological differentiator that operates at normal incidence and in transmission mode. $\textbf{e}_{in}$ and $\textbf{e}_{out}$ are the left and right circular polarizations, respectively. The dielectric slab has refractive index $n = 3.4$ and thickness $a$. 
(b) Multilayer differentiator structure consisting of 16 periods. 
Each period consists of 2 layers: $n_1=3.8$ with thickness $d_1=0.17a$ and $n_2 = 1.47$ with thickness $d_2 = 0.83a$, where $a$ is the length of the period.}
\label{setup}
\end{figure}

In our setup, we consider a left circularly-polarized paraxial beam that is transmitted through the optical differentiator at normal incidence. After transmission, the orthogonal, right circular polarization is selected to form the differentiated output beam. The differentiator consists of either a dielectric slab or a multilayer structure, as shown in Figure \ref{setup}. We define the $+z$ direction as the propagation direction of the incident beam. The transverse $x$ and $y$ axes form a plane parallel to the plane of the differentiator. 

From Fourier optics, the incident transverse electric field $\textbf{E}(x,y)$ can be decomposed into functions of wavevector components $k_x$ and $k_y$ \cite{goodman}: 

\begin{equation}
    \textbf{E}(x,y) = \textbf{e}_{in}\iint \hat{E}(k_x, k_y) e^{ik_xx}e^{ik_yy}dk_x dk_y
\end{equation}
where $\hat{E}(k_x, k_y)$ is the Fourier transform of the scalar field $E(x,y)$ and $\textbf{e}_{in}$ is the input polarization in $x,y$ coordinates. The wavevector $(k_x, k_y) = (0,0)$ corresponds to the normal direction. 

Since transmission through the differentiator is polarization dependent, $\textbf{e}_{in}$ can be expressed in the basis of $s$ and $p$-polarized waves. The optical transfer function, which relates the frequency components of the input and output fields \cite{Ghatak1978}, is given by: $t(k_x, k_y)=\textbf{e}_{out}^\dag M^{-1} T(k_x, k_y)M\textbf{e}_{in}$
where $T(k_x, k_y)$ is the matrix of transmission coefficients for the $s$ and $p$ polarizations $\bigg(\begin{smallmatrix} 
                    t_s(k_x, k_y) & 0\\
                    0 & t_p(k_x, k_y) 
                    \end{smallmatrix} \bigg)$. $M$ is the matrix converting from the $x,y$ basis to the $s,p$ basis: 


\begin{equation}
M = 
\begin{pmatrix}
-\sin{\varphi} & \cos{\varphi} \\
\cos{\varphi} (1+\frac{\theta^2}{2})& \sin{\varphi}(1+\frac{\theta^2}{2})\\
\end{pmatrix}	
\end{equation}
where $\theta=\sin^{-1}(\sqrt{(k_x^2+k_y^2)}/k_0) \approx \sqrt{(k_x^2+k_y^2)}/k_0$ and $\varphi = \tan^{-1}(k_y/k_x)$, as defined in Figure \ref{setup}a. 

For any structure that has continuous rotational symmetry in the $x, y$ plane, the transfer function is independent of the azimuthal angle $\varphi$. Thus, $T(k_x, k_y)$ is an even function with respect to $k_x$ and $k_y$. In the paraxial regime, the lowest order Taylor expansions of the transmission coefficients are: $t_s = t_0 + C_s \theta^2$ and $t_p = t_0 + C_p \theta^2$ where $t_0$ is the transmission coefficient at normal incidence and $C_s, C_p$ are the complex, quadratic coefficients of $t_s$ and $t_p$, respectively.

To operate as a differentiator, the transfer function must satisfy $t(0,0) = 0$, which is achieved by the cross-polarization condition \cite{tengfeng_2021}. 
Once this condition is satisfied, the transfer function for a general input polarization $\textbf{e}_{in}=(e_{in}^x, e_{in}^y)^T$ and output polarization $\textbf{e}_{out}=(e_{out}^x, e_{out}^y)^T$ is: 
\begin{align}
    t(k_x, k_y) &= e_{out}^{x*}e_{in}^x \theta^2 (C_s \sin^2{\varphi} + C_p \cos^2{\varphi}) \nonumber \\ 
    &+ e_{out}^{x*}e_{in}^y \theta^2 \left( \frac{\sin{2\varphi}}{2}(C_p-C_s) \right)  \\
    & + e_{out}^{y*}e_{in}^x \theta^2 \left( \frac{\sin{2\varphi}}{2}(C_p-C_s) \right) \nonumber \\
    &+ e_{out}^{y*}e_{in}^y \theta^2 (C_s \cos^2{\varphi} + C_p \sin^2{\varphi}) \nonumber
\end{align}
Upon transmission through the differentiator, the output electric field is:
\begin{equation}
    \textbf{E}(x,y) = \textbf{e}_{out} \iint \hat{E}(k_x, k_y)t(k_x, k_y) e^{ik_xx}e^{ik_yy}dk_x dk_y
\end{equation}
In our setup, $\textbf{e}_{in}$ is left circularly-polarized and $\textbf{e}_{out}$ is right circularly-polarized, which gives the following transfer function:
\begin{align} \label{transf_func_circ}
    t(k_x, k_y) &= \theta^2 e^{i2\varphi} e^{i \delta} \frac{(C_p - C_s)}{2} = D(k_x + ik_y)^2
 \end{align}
where $\delta$ is a constant phase acquired from wave propagation through the material and $D$ defines the quadratic coefficient of such a differentiator. 
From Equation \ref{transf_func_circ}, we see that the transfer function exhibits a nontrivial topological charge of 2, since $t(k_x, k_y) \propto e^{i2\varphi}$. 

We emphasize that this charge is protected by the symmetry of our structure. 
To show the connection to rotational symmetry, let $H(\textbf{k}) = M^{-1} T(\textbf{k})M$ where $\textbf{k}$ is the wavevector in the $x,y$ plane. We represent a counterclockwise, in-plane $\phi$ rotation 
%
by operators $\widetilde{R}$ and $R$, which operate on the Jones vector space and $\textbf{k}$ space, respectively. Due to the rotational invariance of our system, we have:
\begin{align} \label{rotation_symm_cond}
    \textbf{e}_{out}^\dag \widetilde{R}H(\textbf{k})\widetilde{R}^{-1}\textbf{e}_{in} = \textbf{e}_{out}^\dag H(R(\textbf{k})) \textbf{e}_{in}
\end{align}
When $\textbf{e}_{in}$ is left circularly-polarized, $\widetilde{R}^{-1}\textbf{e}_{in}= e^{i\phi}\textbf{e}_{in}$. Similarly, when $\textbf{e}_{out}$ is right circularly-polarized, $\textbf{e}_{out}^\dag \widetilde{R}=(\widetilde{R}^{-1}\textbf{e}_{out})^\dag=e^{i\phi}\textbf{e}_{out}^\dag$. Simplifying Equation \ref{rotation_symm_cond}, we have:
\begin{align}
    e^{i2\phi}\textbf{e}_{out}^\dag H(\textbf{k}) \textbf{e}_{in} = \textbf{e}_{out}^\dag H(R(\textbf{k})) \textbf{e}_{in}
\end{align}
Thus, $t(R(\textbf{k})) = e^{i2\phi} t(\textbf{k})$ for all $\textbf{k}$ and rotation angle $\phi$, enforced by rotational symmetry. By exchanging $\textbf{e}_{in}$ and $\textbf{e}_{out}$, we can similarly achieve a topological charge of $-2$.
%
%

\begin{figure}[ht!]
\centering\includegraphics[width=\linewidth]{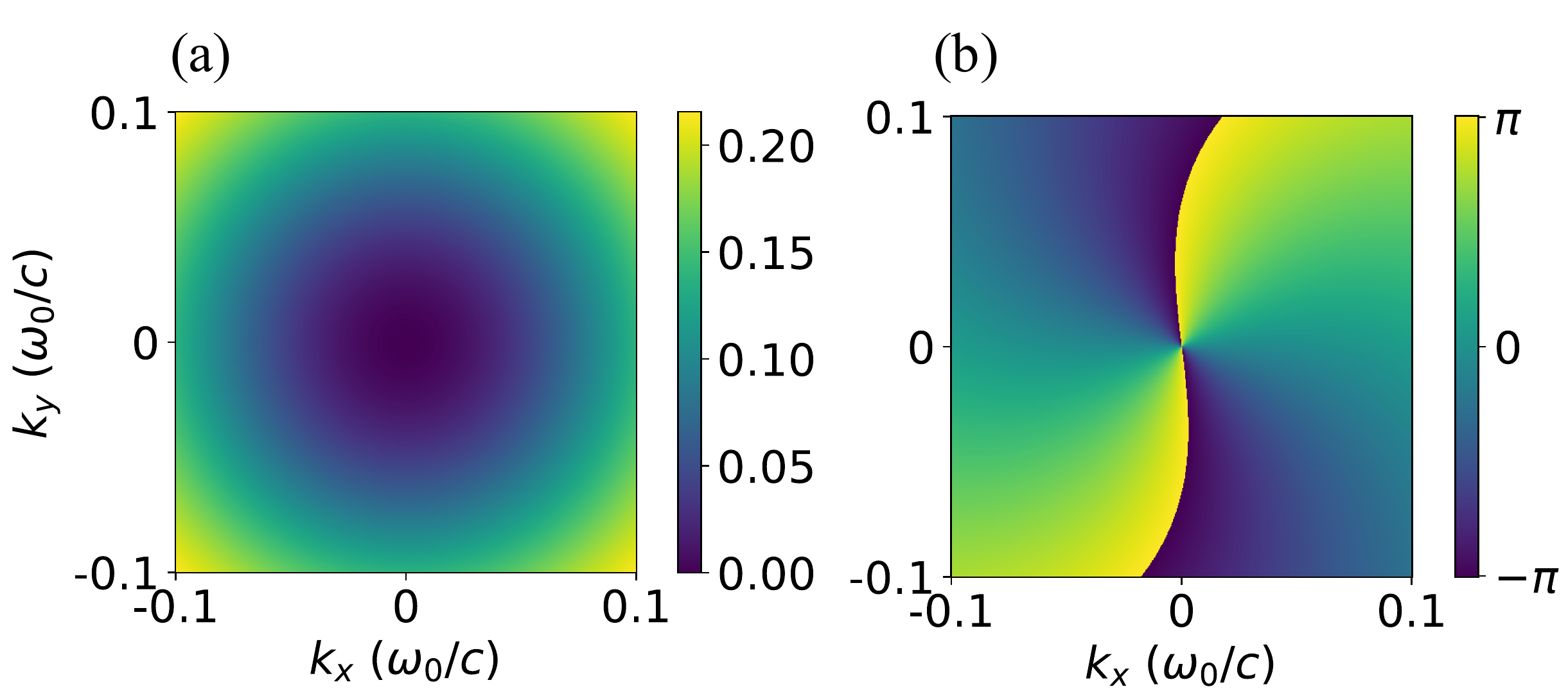}
\caption{(a) Calculated magnitude and (b) phase of transfer function for the multilayer structure with operating frequency $\omega_0=0.188 \times 2 \pi c/a $. 
The structural parameters are specified in Figure \ref{setup}b.
}
\label{multilayer 532 transfer func plots}
\end{figure}

The discussion above is applicable to any structure consisting of one or several uniform dielectric layers. As numerical confirmation of this analysis, we consider the multilayer structure shown in Figure \ref{setup}b. In Figure \ref{multilayer 532 transfer func plots}, we plot the magnitude and phase of the transfer function as a function of the in-plane wavevector, using the operating frequency $\omega_0=0.188 \times 2 \pi c/a $.
As shown in Figure \ref{multilayer 532 transfer func plots}a, the transfer function magnitude only exhibits radial dependence in the $k_x, k_y$ plane, indicating that isotropic differentiation is achieved. Moreover, in Figure \ref{multilayer 532 transfer func plots}b, we observe an isolated zero at the origin with a winding number of 2, since $\frac{1}{2\pi i} \oint_C \frac{dt}{t} = 2$ for any closed curve $C$ encircling the origin and traversed counterclockwise in the $k_x, k_y$ plane.

\begin{figure}[ht!]
\centering\includegraphics[width=\linewidth]{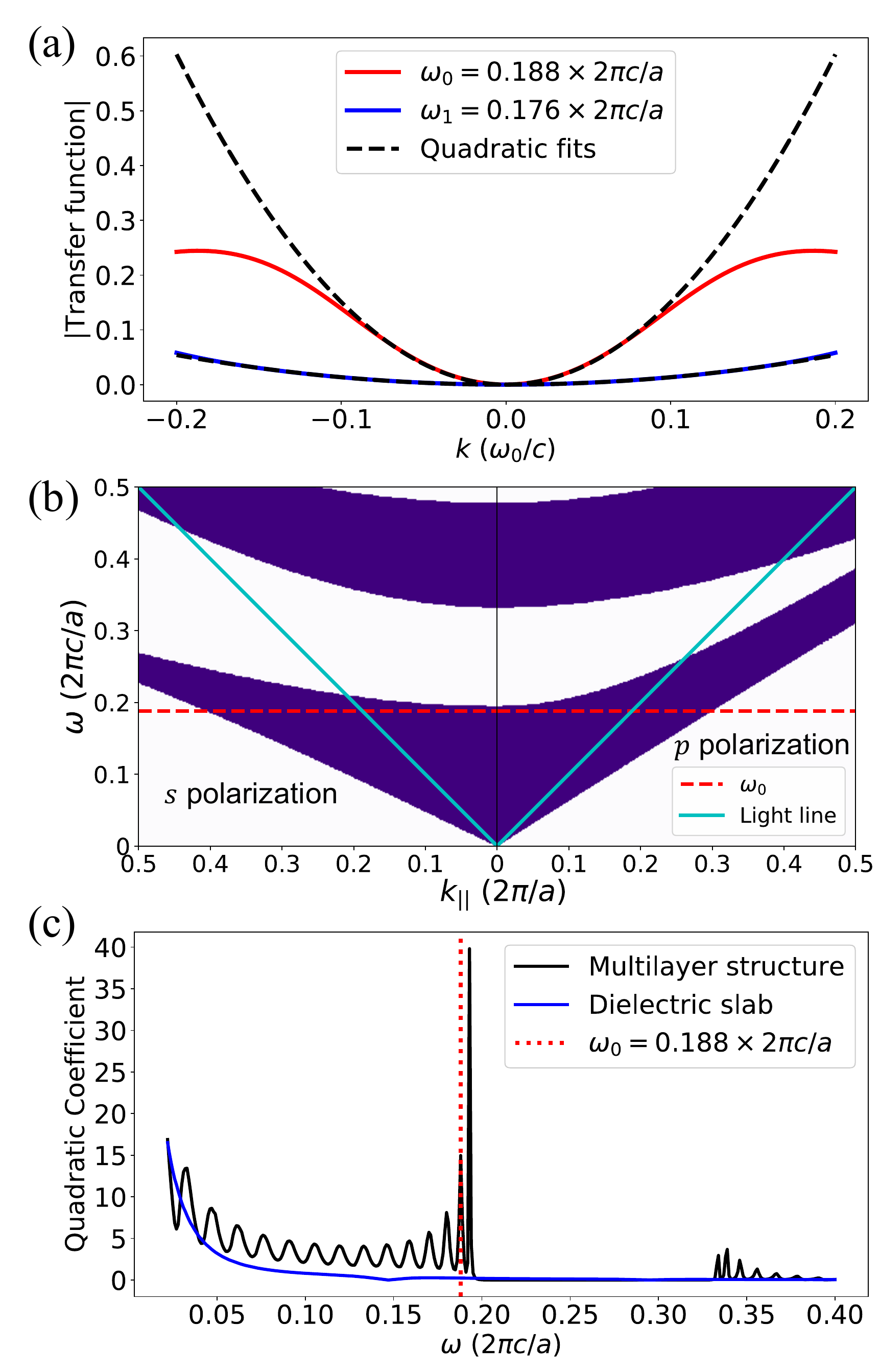}
\caption{(a) Transfer function magnitudes as functions of in-plane wavevector $k$ for selected frequencies $\omega_0=0.188 \times 2 \pi c/a $ (red line) and $\omega_1=0.176 \times 2 \pi c/a $ (blue line).
The dashed black lines indicate the quadratic fits for the $\omega_0$ and $\omega_1$ curves. (b) Projected band diagram of the infinite photonic crystal with the same unit cell as the multilayer structure, with $p$ and $s$ polarizations plotted on the right and left, respectively. The shaded region corresponds to propagating modes and the white region is the photonic band gap. The operating frequency $\omega_0=0.188 \times 2 \pi c/a $ is indicated by the dashed red line, and the blue line marks the light line. (c) Fitted quadratic coefficient of the transfer function magnitude for the finite-sized multilayer structure (black line) and dielectric slab (blue line) as a function of operating frequency. The dotted red line indicates the selected operating frequency $\omega_0=0.188 \times 2 \pi c/a $. }
\label{frequency_sweep}
\end{figure}

Since the multilayer structure has the required rotational symmetry, the functional form of the transfer function in terms of the in-plane wavevector (Figure \ref{multilayer 532 transfer func plots}) remains the same for different operating frequencies. However, the quadratic coefficient $D$, as defined in Equation \ref{transf_func_circ}, varies as a function of frequency. As a case in point, we compare the magnitudes of the transfer functions for frequencies $\omega_0=0.188 \times 2 \pi c/a $ and $\omega_1=0.176 \times 2 \pi c/a $ along the $(k_x, k_y=0)$ direction in Figure \ref{frequency_sweep}a. For both frequencies, the magnitude varies quadratically as a function of $k$ near $k = 0$, but the curvatures of the transfer functions are very different. 

The frequency dependency of the quadratic coefficient can be understood by analyzing the photonic band structure of the multilayer structure. In Figure \ref{frequency_sweep}b, we consider an infinite crystal with the same unit cell as the multilayer film and plot the eigenfrequencies as a function of the in-plane wavevector for both $s$ and $p$ polarizations \cite{Winn98}. At $k = 0$, which describes normally incident light, there is a photonic band gap in the frequency range of $\omega \in [0.195, 0.331] \times 2\pi c/a$. In Figure \ref{frequency_sweep}c, we plot the quadratic coefficient as a function of frequency for the finite-sized multilayer structure (black line). In the frequency range outside the photonic band gap, the coefficient oscillates due to the Fabry-P\'erot resonances in a finite photonic crystal and 
is very strongly enhanced near the edge of the photonic band. In Figure \ref{frequency_sweep}c, we also compare the quadratic coefficient of the 
photonic crystal with that of the uniform dielectric slab from Figure \ref{setup}a. 
We observe significant enhancement of the 
coefficient in the photonic crystal.  

In applying our structure for image differentiation purposes, both the quadratic coefficient and the wavevector range over which the transfer function remains quadratic are important. The magnitude of the quadratic coefficient determines the strength of the differentiation signal, whereas the wavevector range determines the spatial resolution of the differentiator. Moreover, there is generally a trade-off between these two parameters. 
As illustrated in Figure \ref{frequency_sweep}a, the operating frequency $\omega_0 = 0.188 \times 2\pi c/a$ exhibits a much larger quadratic coefficient, but smaller wavevector range over which the transfer function remains quadratic, than the frequency $\omega_1 = 0.176 \times 2\pi c/a$.
In what follows, we will use the operating frequency $\omega_0 = 0.188 \times 2\pi c/a$ to numerically demonstrate edge detection. At this frequency, the quadratic fit works well in the wavevector range of $[-0.07, 0.07] \times\omega_0/c$.
We note that this frequency corresponds to the second peak away from the photonic band edge (Figure \ref{frequency_sweep}c), selected to take into account the trade-off discussed above. Although the first peak away from the band edge exhibits an even larger quadratic coefficient, its wavevector range is too small for differentiation purposes. 

\begin{figure}[ht!]
\centering
\includegraphics[width=\linewidth]{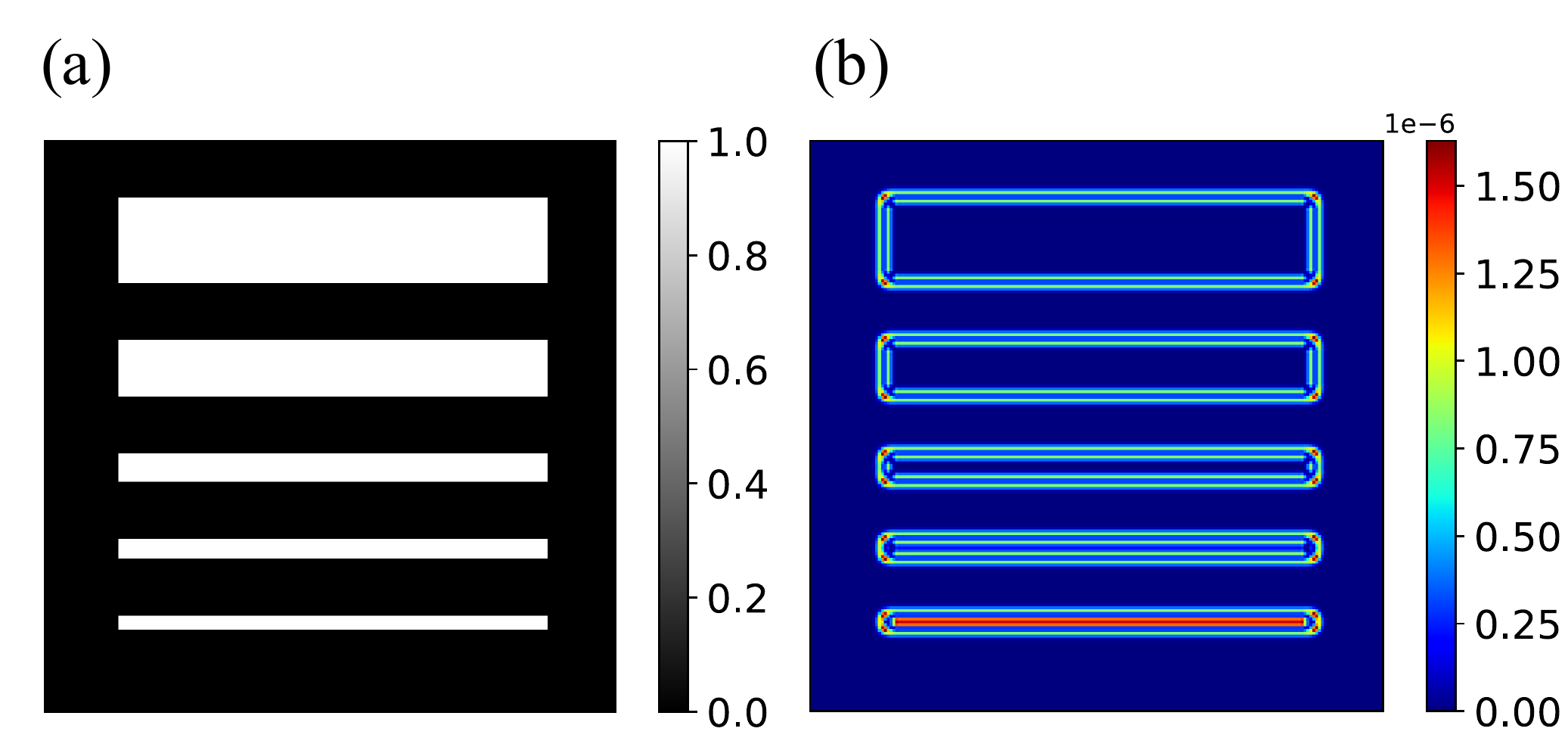}
\caption{Spatial resolution test of differentiator. (a) Intensity of incident image of slot patterns with widths $182a$, $121a$, $60.5a$, $42a$, and $30a$ 
at the operating frequency $\omega_0=0.188 \times 2 \pi c/a $. (b) Intensity of image transmitted through multilayer differentiator.}
\label{resolution test}
\end{figure}

\begin{figure}[h!]
\centering
\includegraphics[width=\linewidth]{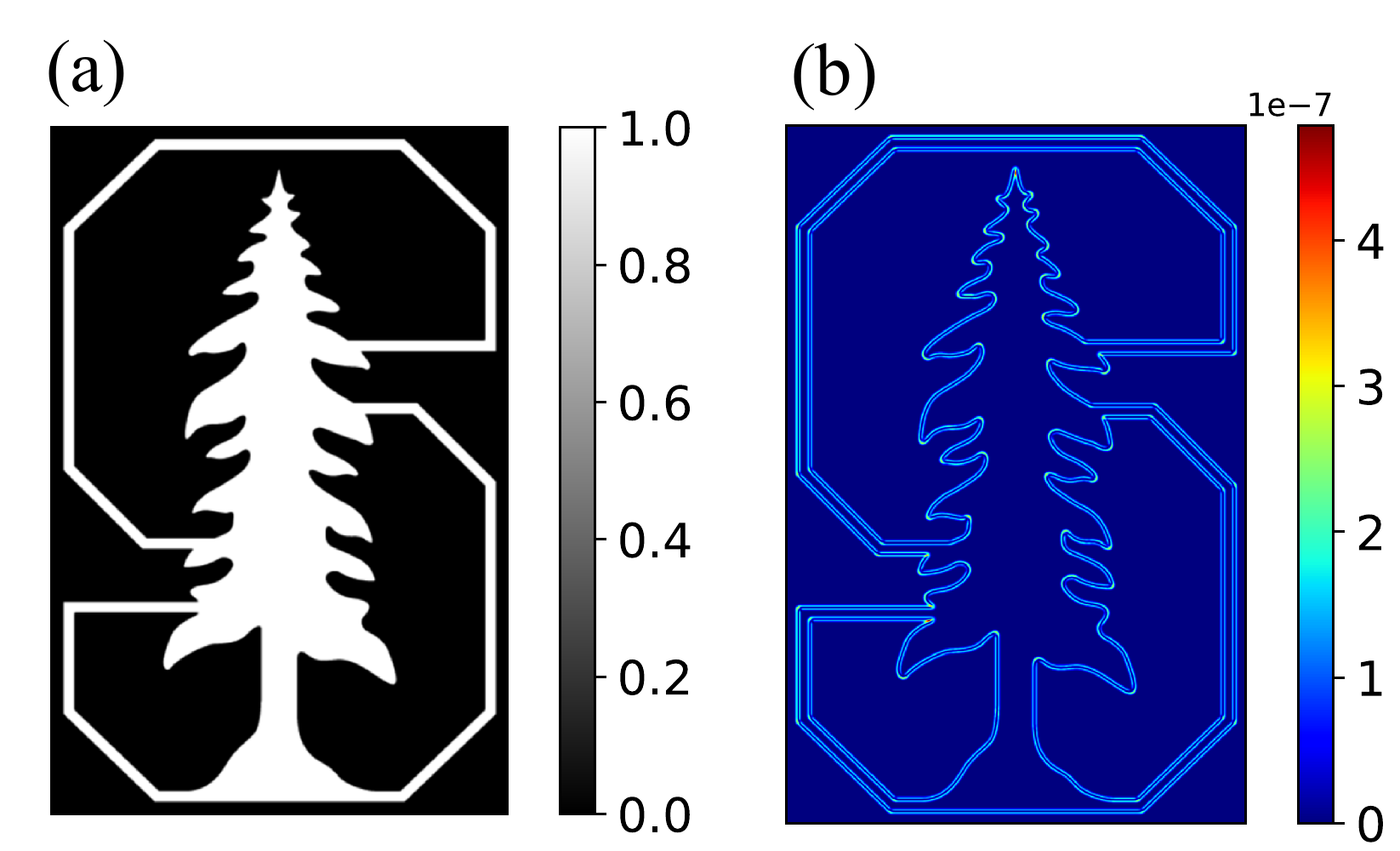}
\caption{Numerical demonstration of isotropic edge detection. (a) Intensity of incident $5970a \times 3950a$ 
Stanford logo at the operating frequency $\omega_0=0.188 \times 2 \pi c/a $. (b) Intensity of image transmitted through multilayer differentiator. }
\label{stanford logo}
\end{figure}

Using our selected operating frequency $\omega_0$ 
for incident $k \in [-0.07, 0.07] \times \omega_0/c$, 
we apply our multilayer differentiator 
to the input images shown in Figures \ref{resolution test} and \ref{stanford logo}. We note that this $k$-range corresponds to a numerical aperture of 0.07 or maximum incident angle of $\theta \approx 4.02^\circ$. During preprocessing, we performed image smoothing by applying a $7\times 7$ and a $11\times 11$
Gaussian kernel to Figures \ref{resolution test} and \ref{stanford logo}, respectively \cite{marr1980theory, shrivakshan2012_gaussian_laplacian}. To test the spatial resolution of our differentiator, we use a series of slot patterns as the incident image, as shown in Figure \ref{resolution test}a. The slot widths correspond to $182a$, $121a$, $60.5a$, $42a$, and $30a$. 
In Figure \ref{resolution test}b, we observe that our differentiator clearly detects the edges of the input image, with a spatial resolution of about $40a$. 
Moreover, each edge generates double lines in the differentiated image, which correspond to the zero-crossings of the second derivative \cite{marr1980theory}.
%
Figure \ref{stanford logo}a shows the $5970a \times 3950a$ 
Stanford logo used as the incident image, which exhibits edges along all directions.
The differentiated image in Figure \ref{stanford logo}b shows that the signal strength is independent of edge orientation
, confirming isotropic differentiation. 


\section{Conclusion}
In summary, we have designed an isotropic, topological differentiator that operates in transmission mode at normal incidence. For circularly polarized input and output light, the transfer function is isotropic and exhibits a topological charge of $\pm 2$ in the $k_x, k_y$ plane. This higher-order charge is protected by the rotational symmetry of our structure, allowing us to achieve second-order differentiation that is robust against imperfections in design parameters and environmental perturbations. 
We have also shown that by using a photonic crystal, and by operating near a band edge of a crystal, we can significantly enhance the differentiation signal.
Furthermore, our design is simpler to fabricate than many of the previously proposed differentiators, which often require surface patterning \cite{dong_HCG_diff_APL_2018, Zhou_PNAS_2019, komar2021edge}.
By engineering a higher-order topological charge in the transfer function, our work points to the importance of topological concepts for optical computing purposes.
\section{Acknowledgments}
\begin{acknowledgments}

This work is supported by a MURI project from
the U.S. Air Force Office of Scientific Research (AFOSR) Grant No. FA9550-17-1-0002,
by a U.S. Office of Naval Research (ONR) Grant No. N00014-20-1-2450 and by a
Vannevar Bush Faculty Fellowship from the U. S. Department of Defense (Grant No.
N00014-17-1-3030). O.L. acknowledges support from the NSF Graduate Research Fellowship and the Stanford Graduate Fellowship.
\end{acknowledgments}

\bibliography{refs}

\end{document}